# Interfacing the JQMD and JAM Nuclear Reaction Codes to Geant4

T. Koi, M. Asai, D. H. Wright
*SLAC, Stanford, CA 94025, USA*

K. Niita
*RIST, Tokai, Ibaraki 319-1106, Japan*

Y. Nara
*Univ. Arizona, Tucson, AZ 85721, USA*

K. Amako, T. Sasaki
*KEK, Tsukuba, Ibaraki 305-0801, Japan*

Geant4 is a toolkit for the simulation of the passage of particles through matter. It provides a comprehensive set of tools for geometry, tracking, detector response, run, event and track management, visualization and user interfaces. Geant4 also has an abundant set of physics models that handle the diverse interactions of particles with matter across a wide energy range. However, there are also many well-established reaction codes currently used in the same fields where Geant4 is applied. In order to take advantage of these codes, we began to investigate their use from within the framework of Geant4. The first codes chosen for this investigation were the Jaeri Quantum Molecular Dynamics (JQMD) and Jet AA Microscopic Transportation Model (JAM) codes. JQMD is a QMD model code which is widely used to analyze various aspects of heavy ion reactions. JAM is a hadronic cascade model code which explicitly treats all established hadronic states, including resonances with explicit spin and isospin, as well as their anti-particles. We successfully developed interfaces between these codes and Geant4. These allow a user to construct a detector using the powerful material and geometrical capabilities of Geant4, while at the same time implementing nuclear reactions handled by the JQMD and JAM models and the Hadronic framework of Geant4 proved its flexibility and expandability.

## 1. INTRODUCTION

As the size and complexity of software has increased, Object-Oriented Programming has become more popular over the last decade. Concurrently, programming languages suitable to the Object-Oriented method have come to be used. C++ is one of them. This trend in software development has also appeared in the simulation of high energy physics experiments. Geant4[1] is one such example. Geant4 is a toolkit for the Monte Carlo simulation of the passage of particles through matter. It succeeds Geant3, which was developed over ten years beginning in 1982 and was written mainly in FORTRAN. Geant4 takes the Object-Oriented Programming approach and uses C++ as the programming language.

Still, there are many fields, such as the simulation of nuclear reactions, in which FORTRAN is the main programming language. Many well-established reaction codes are written and maintained in FORTRAN, as are newly developed codes. While the Geant4 toolkit has its own physics processes, including nuclear reactions, it would be beneficial to users if the FORTRAN-based reaction codes could be made available directly from Geant4. This work investigates the development of interfaces that make this possible. This approach has several advantages over current activities focusing on the translation of existing FORTRAN reaction codes to C++. It is more convenient to make an interface than to re-write an entire code. Also, in the process of re-writing, new bugs may enter into the code. Re-writing the code, on the other hand, provides the chance to uncover bugs in the FORTRAN code. If the reaction code is well established, however, the discovery of many bugs is not likely. Once an interface has been established, the FORTRAN code and Geant4 may be updated independently, as long as developers on both sides adhere to the interface conventions. This is an important feature, not only for the development of Geant4 but also for the FORTRAN-based reaction codes. In addition, no copyright problems associated with re-writes are introduced by the interface methods.

For the first stage of this investigation, two reaction codes, Jaeri Quantum Molecular Dynamics (JQMD) [2] and Jet AA Microscopic Transport Model (JAM)[3], were chosen. Because these codes are well modularized, they are appropriate candidates for the hadronic framework of Geant4 [4]. Also, both codes treat nucleus-nucleus interactions which are much in demand for the simulation of heavy-ion interactions.

The result of the interface development is presented in this paper. A description of the two reaction codes is provided in the next section, followed by an explanation of how these reaction codes were connected to the hadronic framework of Geant4. In section 4 is a discussion of the computing platforms used to develop and test the interfaces. The results of the validation of the interfaced code against heavy-ion interaction data is also presented and discussed.

## 2. REACTION CODE

### 2.1. JQMD

The Quantum Molecular Dynamics (QMD) model is a quantum extension of the classical molecular-dynamics model. The QMD model is widely used to analyze various aspects of heavy ion reactions. The Jaeri QMD (JQMD) model is a QMD code which can also be applied to nucleon-induced and meson-induced reactions. In the JQMD code, a reaction progresses by two steps. First, the direct reactions, non-equilibrium reactions, and dynamical formation of highly excited fragments are calculated in the manner of QMD. Then, evaporation and fission decays are performed for the excited nucleons produced in the first step. This second step is a statistical process and is calculated by a statistical decay model (SDM). Thus JQMD consists of QMD plus SDM. The energy range over which JQMD applies begins at several tens of MeV/N and extends to about 3 GeV/N. It produces the final state particles of the reaction including fragmented nuclei.

A detailed description of JQMD is available in Ref. [2] and the source code of JQMD is available for download from "http://hadron31.tokai.jaeri.go.jp/jqmd/". JQMD is also incorporated in the Particle and Heavy Ion Transport code System (PHITS) [5].

## 2.2. JAM

The Jet AA Microscopic Transportation Model (JAM) is a hadronic cascade model.

In the JAM, the trajectories of all hadrons, resonances and produced particles are followed explicitly in space and time. Inelastic hadron-hadron collisions are modeled by resonance formation and decay at low energies (below ~4GeV). Above the resonance region, string formation and fragmentation into hadrons is assumed. Multiple minijet production is also included at high energies in the same way as HIJING (Heavy Ion Jet INteraction Generator)[6]. Hard parton-parton scattering with initial and final state radiation are simulated using PYTHIA[7]. Rescattering of hadrons which have original constituent quarks can occur with other hadrons assuming the additive quark cross section within a formation time. The effective energy range of JAM extends from several hundreds of MeV/N up to about 100 GeV/N. JAM produces final state particles of the reactions; unlike JQMD, however, it does not produce fragmented nuclei.

A detailed description of JAM is available in Ref. [3] and the source code of JAM can be downloaded from "http://quark.phy.bnl.gov/~ynara/jam/".

## 3. INTERFACING REACTION CODES TO HADRONIC FRAMEWORK OF GEANT4

Before discussing the JQMD and JAM interfaces, it is useful to describe how Geant4 handles hadronic interactions.

The hadronic framework of Geant4 can be divided into two parts. One part manages when and where a hadronic interaction will occur. The generic Geant4 name for the method which does this is GetPhysicalInteractionLength(). The other part deals with the generation of the final state particles of the interaction. This is accomplished in the DoIt() method which, for the purposes of the present investigation, will be implemented by the JQMD and JAM models.

To decide when and where interactions will occur, cross sections are required. In this context, "cross sections" refer not only to scattering cross sections but also to mean life times of unstable particles. However, the decay of unstable particles in Geant4 is beyond the scope of this paper, so "cross sections" will refer here only to inelastic scatterings. However, the decay of excited particles is dealt with in the reaction codes, because the time scale of the decay of resonance particles and strings formed during inelastic reactions is much shorter than the decay of unstable particles in Geant4.

Geant4 has its own cross section tables for many particle species over a wide range of energies. Here, we only refer to cross sections for the inelastic interaction of heavy ions. For heavy ions, Geant4 also has cross sections based on the Triphathi formula[8]. This formula calculates the total reaction cross sections of heavy ions for any system of colliding nuclei and is valid over the energy range from a few A MeV to a few A GeV. The Triphathi formula is an empirical formula, however the parameters are associated with the physics of the collision system. Heavy-ion cross sections based on the Shen formula[9] have also been prepared. This is also an empirical formula, however with a different form and parameters. Users may select which formula to use for the calculation of heavy-ion cross sections. For both formulae, the agreement between the calculated and experimental data is good.

The integration of JQMD and JAM interfaces into the Geant4 hadronic framework is shown schematically in Fig. 1. The inelastic interaction of heavy ions is handled by the G4HadronicProcess class, which contains a pointer to the G4HadronicInteraction class. The concrete models JQMD2G4InelasticModel and JAM2G4InelasticModel are derived from G4HadronicInteraction, and contain the interfaces to JQMD and JAM, respectively. The JQMD or JAM FORTRAN subroutine responsible for generating the final state particles is called from the ApplyYourSelf() method of the corresponding model class. The cross sections are handled by the G4HadronicProcess class.

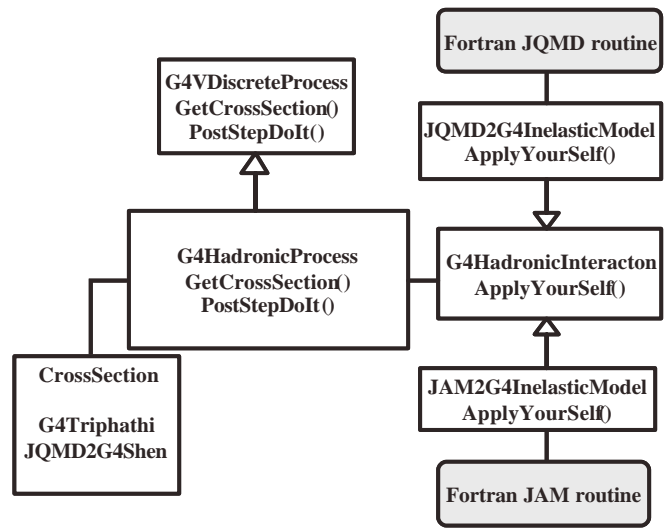

Figure1, A schematic view of how FORTRAN reaction codes are interfaced to the Geant4 hadronic code for heavy-ion interactions.

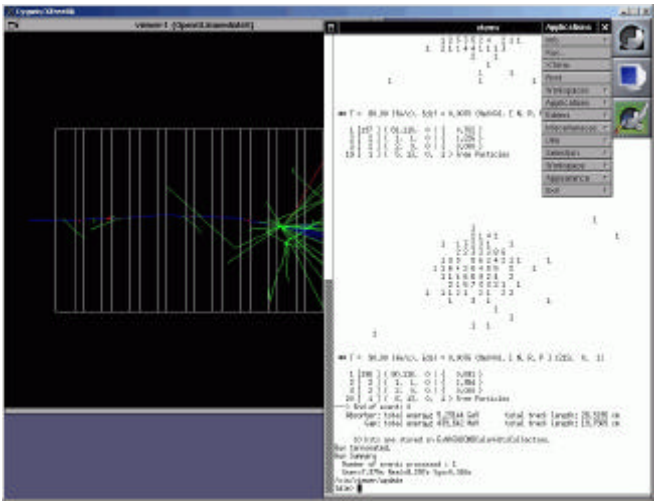

Figure 2, Screen shot of the JQMD interface in use with Geant4.

## 4. DEVELOPMENT OF INTERFACE AND RESULT

The present investigation was carried out on the following platforms:

OS: Red Hat Linux 7.2 and also 6.2
Compiler: gcc-2.95.3 and also *2.91.66 with egcs-1.1.2*
Geant4: Ver. 5.0.p01 (2003 Feb version).

The interfaces were successfully developed and run. Fig. 2 is a screen shot of the JQMD interface in use. In the case shown, an iron ion enters a sandwich calorimeter which is described in novice example N03 of Geant4. The right side of the screen shows the nucleon distribution in the QMD calculation. This screen shot was obtained on Cygwin 1.3.22-1 with gcc 3.2.3.

The double differential cross sections of pion production from nucleus-nucleus interactions were predicted using the developed interfaces. One of the results is shown in Fig. 3 where it is compared with experimental data[10]. This figure shows the energy spectrum of pions produced by 3.0 GeV/A alpha particles on a copper target. The pions were emitted at an angle of 5 degrees with respect to the incident alpha direction. The results from JQMD and JAM agree well with one another except at low momenta. Both models reproduce the data well.

These interfaces were not tested on other operating systems and compilers. It is likely that only small modifications will be required for these interfaces to work on other platforms supported by Geant4.

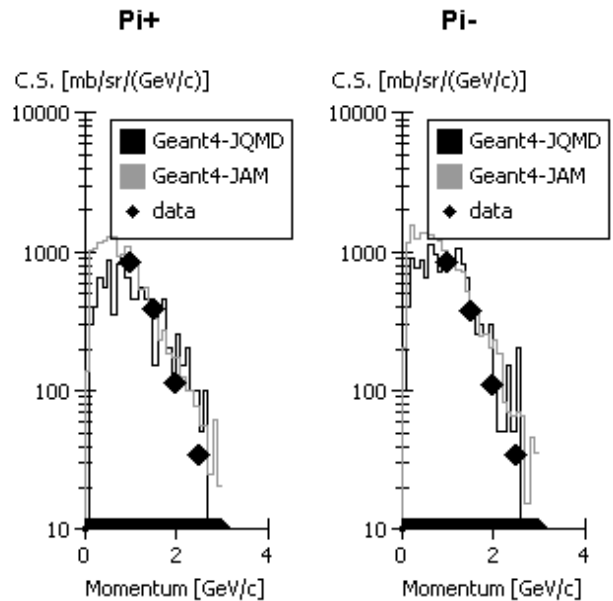

Figure 3, Pion production from alpha particles bombarding a copper target. Positive pions are shown on the left and negative pions on the right.

## 5. CONCLUSION

Interfaces to connect JQMD and JAM to Geant4 have been successfully developed. The interfaces have been validated against several experimental data sets. Even though these data sets were well-reproduced, further validation efforts are required. The success of this investigation indicates the flexibility and expandability of the Geant4 hadronic framework. With this interface, the simulation of heavy-ion propagation in complex Geant4 geometries is now possible. There are no special limitations imposed by the use of these interfaces. They are therefore available to all Geant4 users who wish to use them.

Recently, the demand for heavy ion simulations has increased from people who are working not only in high energy and nuclear physics but also in cancer therapy, space applications and many other engineering and science fields. These interfaces will be of use in all these fields.

### Acknowledges

The authors wish to thank Dr. S. Chiba at JAERI for his efforts in making the JQMD code publicly available. The work of T. K., M. A. and D. H. W. (SLAC) was supported by Department of Energy contract DE-AC03-76SF00515.